\documentstyle[prcrc,11pt,fleqn]{article}

\title{Primordial black holes, phase transitions, and the fate of the
universe\footnote{Talk presented by the first author at the Primordial
Black Hole Workshop/Dark Matter 1998, held in Marina del Rey, CA, from
February 17-20, 1998.}}

\author{Mitesh Patel and George
M. Fuller\address{Department of Physics, University of California, San
Diego, La Jolla, California 92093-0350}\footnote{Email addresses:
{\tt mitesh@physics.ucsd.edu}, {\tt gfuller@ucsd.edu}.}}

\begin{document}
\maketitle

\begin{abstract}
Phase transitions in the early universe are prime settings for the
production of primordial black holes, since they can break the
relatively quiescent homogeneity and isotropy of
Friedmann--Robertson--Walker (FRW) cosmologies. These epochs of
``symmetry breaking,'' moreover, can affect the subsequent development
of spacetime by changing the evolution of some FRW parameters,
including the present age and density of the universe. We discuss the
relative importance of such effects on constraining mechanisms of
black hole formation.
\end{abstract}

\section{Introduction}
Although they have not yet been observed, primordial black holes
(PBHs) already deserve a special place in the temple of modern
theoretical physics, for they have spawned many creative ideas at the
intersection of cosmology, astrophysics, and particle physics. In
cosmology, for example, they may affect the outcome of Big Bang
nucleosynthesis (BBN) \cite{BBNlimits}. Primordial black holes may be of
astrophysical interest through their evaporative production of the
highest energy cosmic rays \cite{cosmicrays}. Their formation during cosmic
phase transitions (e.g., the electroweak (EW) and quantum chromodynamic
(QCD) transitions) also may help constrain the relevant particle
physics. Further examples of this rich spectrum of applications abound
in these proceedings. We consider below the
example of Massive Compact Halo Object (MACHO) black holes and discuss
two bulk cosmological constraints on the production of such holes.

\section{An example: MACHO black holes formed at the QCD epoch}
One of the recent uses of PBHs involves the dark matter in the
halo of our Galaxy. The MACHO Project
reports that the Galactic halo contains condensed objects of about
0.2 to 0.8 solar masses \cite{griest}. Since candidates such as
ordinary stars, white dwarfs, neutron stars, etc., are constrained by
various observations such as the Hubble Deep Field, stellar and
cosmological nucleosynthesis, etc., we must consider more exotic
possibilities to explain the MACHO events. Primordial black holes
evade these constraints, because they would have formed in the
early universe before BBN.

In order to produce MACHO-sized black holes, the existence of particle
horizons requires that we study an epoch during which a
horizon volume contains 
at least about one solar mass in mass-energy. The early universe is
characterized by a very high degree of homogeneity and isotropy of
spacetime. Further, any isocurvature fluctuations larger than a
horizon length are ``frozen,'' and those smaller are damped. The only
reasonable way to produce PBHs, therefore, must involve the
amplification of pre-existing curvature fluctuations (those laid down
by inflation, for example) or the creation of such perturbations as a
result of phase transitions (PTs). Typical first-order PTs, however,
nucleate bubbles of the broken phase that are small compared to the
horizon, so they cannot make MACHO-sized black holes. Second-order PTs
have even less spectacular consequences, so we focus on the possibility
that a first-order PT amplifies pre-existing fluctations to the point of
gravitational collapse into roughly horizon-sized black
holes. Since the mass in the horizon at the QCD epoch is roughly one
solar mass, there is a chance that some horizon volumes will ``go
down'' into MACHO-sized black holes during or just after the QCD
transition (see, for example, Ref. \cite{jednieful} for detailed
discussions of specific formation mechanisms).

\section{Two constraints on PBH formation}
The mechanism of black hole formation had better not be too efficient,
for
energy density in black holes redshifts like ordinary matter (i.e.,
more slowly than radiation), and copious PBH production would make the
universe prematurely dominated by non-relativistic matter, precluding
crucial cosmological events like BBN. Another way to state this is
that the universe becomes ``overclosed'' (i.e., $\Omega_{\rm PBH}$
today is inconsistent with the present observational bounds on
$\Omega_0$) unless there is at most one PBH formed per $10^7$ horizon
volumes \cite{hall}. This is a powerful constraint on building
models of PBH formation.

Another possible constraint involves the manifest breaking of
Friedmann--Robertson--Walker (FRW) symmetry when a PBH forms. Using the
previous constraint, we can view the universe as a lattice of comoving
black holes with a lattice separation of at least $\approx 215$
horizons at the epoch of formation. The horizon eventually expands to
encompass many holes, at which time they may attract each other and
cluster together. Since black holes are the ultimate curvature
fluctations, spacetime expands non-uniformly, and the overall Hubble
expansion of the universe is valid only in an averaged sense. In
particular, these fluctuations effect a ``back-reaction'' on the
expansion of the universe and may change the values of $\Omega_0$ and
$H_0$ observed today (i.e., the fate of the universe). If the
resulting theoretical values disagree significantly with the current
values, then this is a valid constraint on PBH formation.

The latter constraint is quantitatively calculable using a general
relativistic perturbation theory in a suitable gauge but, with minimal
perspiration, we can conclude that this effect is utterly
negligible. First,
we expect the effect of inhomogeneities on the expansion or age of the
universe to be more significant during the epoch of matter domination
than radiation domination, since the former lasts much longer than the
latter and since nonlinear structures such as galaxies and clusters
may form only during the former epoch. Using a general relativistic
analog of Zel'dovich's pancake approximation for gravitational collapse,
Russ and collaborators \cite{zeldoruss} have calculated the change in the
age of the universe due to growing inhomogeneities. They find that the
age of the universe decreases from the usual FRW value by a part in
$10^3$ to $10^4$, depending on the composition of the dark
matter. Therefore, the potential constraint on PBH production
described in the previous paragraph is {\it very} weak. To put it another
way: the constraint may become important only when we know the age of
the universe to at least three significant figures.

This work was supported in part by grants from the NSF and NASA.


\begin{thebibliography}{9}

\bibitem{BBNlimits} A.~R.~Liddle and A.~M.~Green, these proceedings,
preprint gr-qc/9804034, {\it Phys. Rept.} {\bf 307} (1998) 125.
\bibitem{cosmicrays} J.~H.~MacGibbon and B.~J.~Carr, {\it
Astrophys. J.} {\bf 371} (1991) 447; B.~J.~Carr and J.~H.~MacGibbon,
these proceedings, {\it Phys. Rept.} {\bf 307} (1998) 141.
\bibitem{griest} C.~Alcock, {\it et al.}, {\it Phys. Rev. Lett.} {\bf
74} (1995) 2867; C.~Alcock, {\it et al.}, {\it Astrophys. J.} {\bf
461} (1996) 84; K.~Griest, U.C. San Diego, private communication, 1998.
\bibitem{jednieful} K.~Jedamzik, these proceedings, preprint
astro-ph/9805147; J.~Niemeyer, these proceedings, preprint
astro-ph/9806043; G.~M.~Fuller, these proceedings.
\bibitem{hall} L.~J.~Hall and S.~D.~Hsu, {\it Phys. Rev. Lett.} {\bf
64} (1990) 2848.
\bibitem{zeldoruss} M.~Kasai, {\it Phys. Rev.} {\bf D52} (1995) 5605;
H.~Russ, M.~Morita, M.~Kasai, and G.~B\"orner, {\it Phys. Rev.} {\bf
D53} (1996) 6881; H.~Russ, M.~H.~Soffel, M.~Kasai, and G.~B\"orner,
{\it Phys. Rev.} {\bf D56} (1997) 2044.


\end{thebibliography}
\end{document}